# Superseding traditional indexes by orchestrating learning and geometry

## Giorgio Vinciguerra
Department of Computer Science, University of Pisa, Pisa, Italy
giorgio.vinciguerra@phd.unipi.it
https://orcid.org/0000-0003-0328-7791

## Paolo Ferragina
Department of Computer Science, University of Pisa, Pisa, Italy
paolo.ferragina@unipi.it
https://orcid.org/0000-0003-1353-360X

## Michele Miccinesi
Department of Computer Science, University of Pisa, Pisa, Italy
m.miccinesi@studenti.unipi.it
https://orcid.org/0000-0003-3138-8063

—— **Abstract** ——

The explosion of big data and the new generation of applications and computing paradigms, possibly related to mobile and IoT scenarios, raise new challenges, such as strict latency, energy and storage constraints, which usually vary among devices, users and time because of intrinsic and extrinsic conditions not predictable in advance. The algorithmic community addressed these challenges by focusing, among the others, on memory hierarchy utilisation [15, 38], query processing on streams [11], space efficiency [30, 35], parallel and distributed processing [19], and software auto-tuning [29], just to mention a few.

In this paper, we contribute to these studies in a twofold manner. We design the first learned index (à la [27]) that solves the dictionary problem with provably better asymptotic time and space complexity than classic indexing data structures for hierarchical memories, such as B-trees and Cache-Sensitive Search trees [34], and modern learned indexes [16, 27]. We call our solution the *Piecewise Geometric Model index* (shortly, PGM-index) because it turns the indexing of a sequence of keys into the coverage of a sequence of 2D-points via linear models (i.e. segments) suitably learned to trade, in a principled way, query time vs space efficiency. This idea comes from the heuristic results of [16, 27] which we strengthen by showing that the minimal number of such segments can be computed via known and optimal streaming algorithms [32, 42]. The PGM-index is then obtained by recursively applying this geometric idea and by guaranteeing a smoothed adaptation to the "geometric complexity" of the input data. Finally, we propose a variant of the PGM-index that adapts itself not only to the distribution of the dictionary keys but also to their access frequencies, thus obtaining the *first distribution-aware learned index*.

The second main contribution of this paper is the proposal and study of the concept of *Multicriteria Data Structure*, namely one that adds to the classic requirements of being space and time efficient, the novel feature of being flexible enough to dynamically adapt itself to the constraints imposed by the application of use. We show that the PGM-index is a multicriteria data structure because its significant flexibility in storage and query time can be exploited by a properly designed optimisation algorithm that efficiently finds its best design setting in order to match the input constraints (either in time or in space).

Such theoretical contributions are corroborated by a thorough experimental analysis over three known and large datasets, showing that the PGM-index and its multicriteria variant improve uniformly, over both time and space, classic and learned indexes up to several orders of magnitude. This makes our geometrically-learned approach potentially useful in modern DB-scenarios [26] and paves the way to novel investigations in the classic realm of data structure design for other problems, some of which are stated in the concluding section of this paper.

**2012 ACM Subject Classification** Information systems → Data structures, Theory of computation → Data structures and algorithms for data management, Theory of computation → Data compression

**Funding** Part of this work has been supported by the EU grant for the Research Infrastructure "SoBigData: Social Mining & Big Data Ecosystem" (INFRAIA- 1-2014-2015, agreement #654024) and by a Google Faculty Award 2016 on "Data Compression"

## 1 Introduction

The ever-growing amount of information coming from Web, social networks and Internet of Things seriously slows down the management of available data. Advances in CPUs, GPUs and memories hardly solve this problem without properly devised algorithmic solutions. Hence, much research has been devoted to dealing with this enormous amount of data, particularly focusing on memory hierarchy utilisation [15, 38], query processing on streams [11], space efficiency [30, 35], parallel and distributed processing [19].

Despite the formidable results achieved in these areas, we still miss proper algorithmic solutions that are flexible enough to work under computational constraints that vary across users, devices and time [43, 17]. As an example, fog and edge computing need the orchestration of heterogeneous devices like routers, smartphones, wearables and sensors so diverse in latency, energy and storage constraints that it is too difficult for a software engineer to choose the correct algorithmic solution on a per-device basis [28].

In this paper, we aim at formally digging into these issues, and so we restrict our attention to the case of *indexing data structures* that solve the classical *static dictionary problem*. Classic solutions can be grouped into four main families [22]: (i) hash-based, which range from traditional hash tables to recent techniques, like Cuckoo hashing [33]; (ii) tree-based, such as B-trees and its variants [18, 34]; (iii) bitmap-based [8], which can take advantage of compression techniques; and (iv) trie-based, which are commonly used for string keys. Unfortunately, hash-based indexes do not support predecessor or range searches; bitmap-based indexes can be expensive to store, maintain and decompress [39]; whereas tree- and trie-based indexes are mostly pointer-based and, apart from recent compression results [14], keys are stored uncompressed thus taking space proportional to the dictionary size.

A novel and somewhat surprising approach to dictionary indexing was recently proposed by [27], expanding and improving some previous results of [2]. These authors suggested that indexes can be interpreted as models mapping keys to their rank in the sorted order (i.e. their positions). To better understand this statement, let us sort and store the $n$ keys of the input dictionary $D$ in an array $A$. Then, we can interpret any indexing data structure as a function $index : \mathcal{U} \to [1, n]$ from the universe of keys to a position into $A$. If $D$ contains repeated keys, $index$ returns the location of the first occurrence in $A$ of the queried key (if it exists), so that the records sharing the same key can be retrieved by scanning the following positions in optimal output-sensitive manner. The function $index$ could be easily extended to manage the case of keys not in $D$ by returning either the special value 0 or the position in $A$ of its *predecessor*. Looking from the perspective of a mapping, this interpretation of the $index$ function seems not much a new one. Any one of the previous four families of indexes provides efficient, sometime asymptotically optimal, implementations of $index$. The novelty of the scenario proposed in [27] becomes apparent when we look at the input keys $k \in A$ as points $(k, index(k))$ into the Cartesian plane, as shown in Figure 1. In this case, we can look at the implementation of *index* as a Machine Learning (ML) problem in which we search for the most succinct and efficient model that best approximates that function.

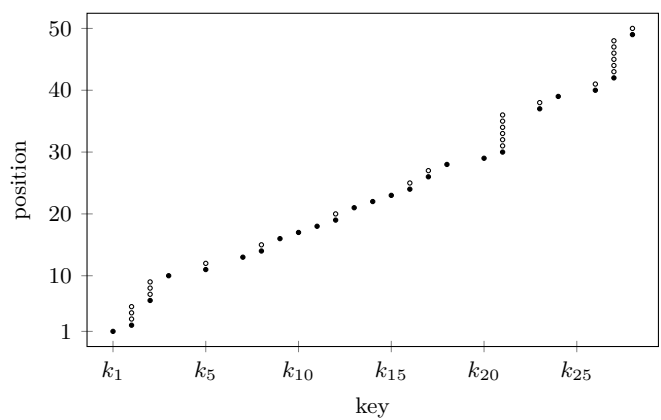

**Figure 1** A dictionary of ordered keys, represented as 2D-points $(k, index(k))$.

To better understand the potentiality of this approach, let us start with a trivial example. Consider the case of a dictionary of integer keys $a, a + 1, a + 2, \dots, a + n - 1$, where $a$ may be any integer. Here, $index(k)$ can be computed easily as $k - a + 1$, and thus it takes constant time and space to be implemented, independently of the number $n$ of keys to be indexed. This example can be generalised to any set of keys $k_i$ whose pairs $(k_i, i)$ are distributed over a line, with a proper *slope* and *intercept*. In this case too, the *index* function for a key $k$ can still be implemented in constant time and space as $index(k) = slope \times k + intercept$, which is again independent of the number $n$ of keys to be indexed. These simple examples shed light on the potential compression opportunities offered by patterns and trends in data distribution which are more frequent than expected, as we will show in the experimental sections. However, we cannot argue that all datasets follow a "linear trend", nor that the algorithm designers develop ad-hoc solutions for each real-world data distribution. In the general setting, we have therefore to design proper ML techniques that *learn* the *index* function by extracting the patterns present in the input data by means of proper models which may range from linear to more sophisticated functions. Clearly, learning has to be orchestrated with *efficiency in query time and occupied space* of the learned model, since *index* must be used on-the-fly in some running applications.

This is exactly the design goal pursued very recently by Kraska et al. [27] with their Recursive Model Index (RMI), which uses a hierarchy of ML models organised as a Directed Acyclic Graph (DAG) and trained to learn the input data distribution, such as the one of Figure 1. At query time each model, starting from the top one, takes the queried key $k$ as input and picks the following model in the DAG that is "responsible" for that key. The output of RMI is the position returned by the last queried ML model, which is, however, an *approximate position* for the queried key in $A$. A final binary search is thus executed within a range of neighbouring positions whose size depends on the prediction error of RMI.

One could presume that ML models cannot provide the guarantees ensured by traditional indexes, both because they can fail to learn the distribution and because they can be expensive to evaluate [26]. Unexpectedly, on datasets of 200M entries, Kraska et al. [27] reported that RMI always dominates the B-tree performance, being up to 1.5–3× faster and, very significantly, two orders of magnitude smaller in space.

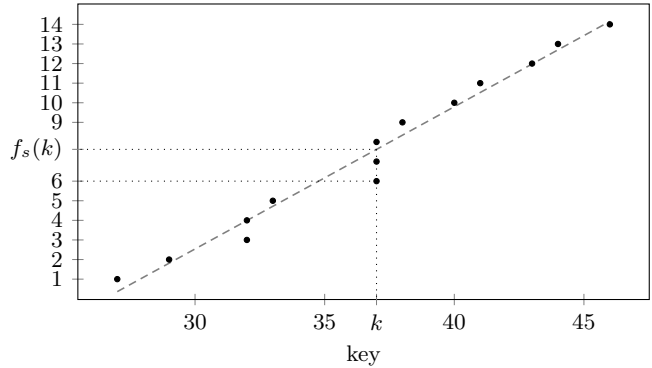

 **Figure 2** An example of linear approximation of a set of integer keys within the range $[27, 46]$. The encoding of the linear model takes only two floats, and thus it is independent of the number of "encoded" keys, provided that they can be "linearly learned" with an $\varepsilon$-error.

This surprising result is stimulating a lot of interest in the algorithmic and the DB communities [4, 22, 26] on RMI's design and limitations. The RMI introduces *another set of* space-time trade-offs between model size and query time which are difficult to control because they depend on the data distribution, on the DAG structure and on the complexity of the ML models adopted. This motivated the recent introduction of the A-tree [16] which uses linear models only, a B-tree structure to index them, and provides an integer parameter $\varepsilon \geq 1$ controlling the size of the region in which the final binary search step has to be performed. Figure 2 shows an example of a linear model $f_s$ approximating fourteen keys within the range $[27, 46]$ and its use in determining the approximate position of a key $k = 37$, which is indeed $f_s(k) \approx 8$ instead of the correct position 6, thus making an error $\varepsilon = 2$. The experiments of [16] show that the A-tree improves the time performance of the $B^+$-tree with a space saving of up to four orders of magnitude. However, the major limitation of A-trees lies in the way they compute the linear approximation of the input keys: the algorithm is sub-optimal in theory and pretty inefficient in practice. This prevents us to fully exploit the space-time potentiality of such learned indexes (as we will show in the experimental sections).

In this paper, we contribute to both the design of *optimal* learned indexes and the *automatic selection of the best* (learned) index that fits the requirements of an underlying application in five main steps, which we summarise below.

1. We orchestrate geometric and learned indexes via a novel *recursive* approach, so to obtain what we call the *Recursive PGM-index*. This index adapts smoothly to the "geometric complexity" of the input keys, for which we also provide some non-trivial lower bounds (see Section 2.1.2). As a result, the PGM-index is the first learned index which is provably better than classic and learned indexing data structures (see Section 2.3, Theorem 6, and Table 1).

2. We propose a variant of the PGM-index that adapts itself not only to the distribution of the dictionary keys but also to their access frequencies. This novel contribution, which we call *Distribution-Aware PGM-index*, has the query time of biased data structures [3, 5, 13, 25, 36], but a space occupancy that does not depend on the number of keys and adapts to the "geometric complexity" of the dataset, thus resulting very succinct in space too (see Section 3 and Theorem 7).

3. We show how to generalise the PGM-index to use not only linear models but also more complex regression models, such as neural networks of few neurons (just to keep their space succinct). In this case, we devise an optimisation strategy that explores the space of regression models to be fit in the structure of the PGM-index. We call this new data structure *Hybrid PGM-index*, and we show that it empowers the hybrid indexes introduced by Kraska et al. [27] with the additional features to be constructed automatically and with a user-controlled approximation quality. The net result is that our PGM-index turns out to be as general as the RMI [27] and with theoretical guarantees better than the A-tree [16], therefore resulting in a replacement for both of them (see Section 2.2).

4. We design a framework that automatically and efficiently optimises the PGM-index given a space or a time constraint. This is where the combination between multicriteria optimisation and learned indexes shows its full potential, and leads us to devise the novel concept of *Multicriteria Data Structure* that we introduce in this paper and which finds in the PGM-index one possible instance (see Section 4). A multicriteria data structure, for a given problem $P$, is defined by a pair $\langle \mathcal{F}, \mathcal{A} \rangle_P$ where $\mathcal{F}$ is a family of data structures, each one solving $P$ with a proper trade-off in the use of some computational resources (such as time, space, energy, etc.), and $\mathcal{A}$ is a properly designed optimisation algorithm that efficiently selects in $\mathcal{F}$ the data structure that "best fits" an instance of $P$. We demonstrate the fruitfulness of this new class of data structures by focusing on the paradigmatic static dictionary problem in which $\mathcal{F}$ is the family of PGM-indexes, and the optimisation algorithm exploits a simple space-time cost model to explore $\mathcal{F}$ efficiently. This result makes the PGM-index the first multicriteria data structure to date.[1]

5. Our last contribution is a thorough set of experiments over some known and large datasets (see Section 5), in which we compare the performance of our PGM-index against classic indexes (namely, B-trees and CSS-trees) and learned indexes (namely, A-trees and RMIs). For example, we show that our index improves the space occupancy of the A-tree by 60%, of the CSS-trees by a factor $82.7\times$, and of the B-tree by more than four orders of magnitude, while achieving their same or even better query efficiency. With respect to the RMI, the PGM-index offers theoretical guarantees on query time and space occupancy, and the experiments on its Multicriteria variant show improved performance. In summary, our experimental results for the Multicriteria PGM-index support the vision of a new generation of big data processing systems, in which data structures and their algorithms can be tailored to the application, device and user [21, 26].

Overall we believe that our results, apart from their individual contributions, pave the way to many other challenging theoretical and algorithm-engineering problems that we summarise in the last Section 6.

## 2    The Piecewise Geometric Model Index

The Piecewise Geometric Model index (PGM-index) is a parameterised data structure which solves the static dictionary problem over a sorted array $A$ of $n$ keys from a universe $\mathcal{U}$ of real numbers. Precisely, given an integer parameter $\varepsilon \geq 1$, the PGM-index learns an approximate mapping between keys from the universe $\mathcal{U}$ and their positions in the input array $A$.[2] Given a query key $k$, the PGM-index finds first a position *pos* in $A$ which is at most $\varepsilon$ away from the correct position of $k$, and then searches $k$ in the subarray $A[pos - \varepsilon, pos + \varepsilon]$ via binary search or other approaches.

---

[1]  For completeness, we mention that the concept of multicriteria optimisation has been already applied in Algorithmics to data compression [12], compiler optimisation [20], and software auto-tuning [29], but as far as we know it is new in data structures design.

[2]  As anticipated in the introduction, if $k \notin A$ the "position" of $k$ is the one of its *predecessor* in $A$.

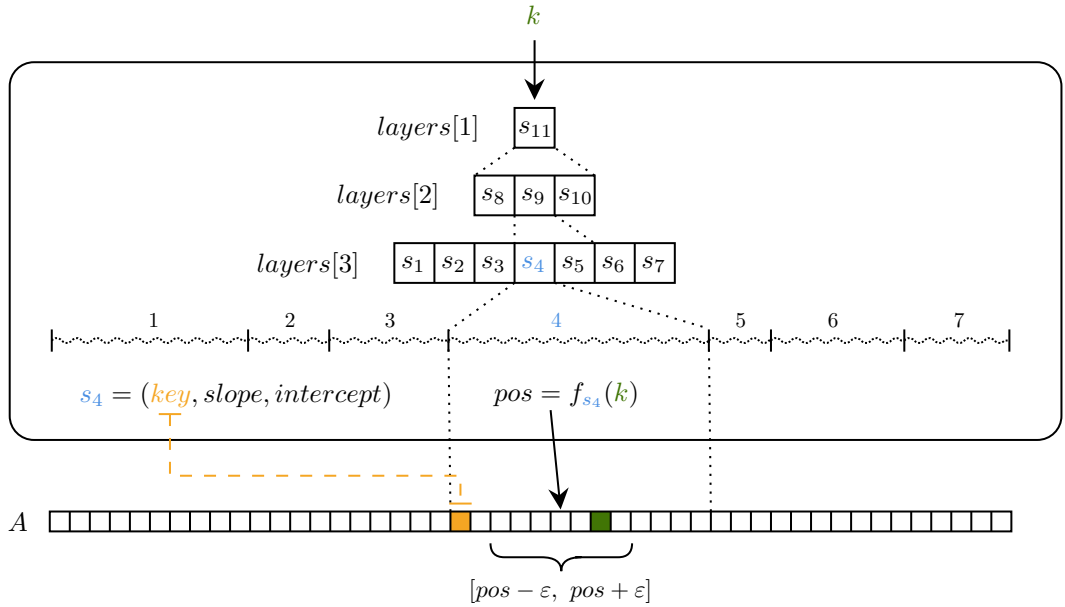

The key tool on which the definition of the PGM-index hinges on is a Piecewise Linear Approximation model (PLA-model) that implements efficiently in time and space the approximate mapping from a set of keys to their positions in the sorted array $A$. By leveraging this tool, described in Section 2.1, the PGM-index turns an array of keys $A$ into an array of linear models (i.e. segments) each one taking a constant space (namely, two floats and one key) and constant query time to return the $\varepsilon$-approximate position of a queried key in the indexed subarray. Therefore, every segment is independent in time and space of the number of its indexed keys. In the rest of this section, we will concentrate the description of the PGM-index to the case of linear models and defer to Section 2.2 its generalisation to the case of nonlinear models which are used to $\varepsilon$-approximate the positions of the keys in $A$.

The PGM-index is built on top of the array of linear models in several ways which are explored in Section 2.3, e.g. via a B-tree or a Cache-Sensitive Search tree (CSS-tree). Actually, any indexing data structure could be adopted on top of the array of models to route the search for a key among them. However, each of these solutions would not make the most of the constant space-time indexing feature offered by a single linear model. As a consequence, we design the overall PGM-index by recursively constructing a series of PLA-model as follows. Firstly, we construct the PLA-model over the whole array $A$. Secondly, we turn the individual models of this PLA-model into a "proper set of keys". Thirdly, we construct another (and smaller) PLA-model over those keys. This process continues recursively until one single linear model is obtained, which will form the root of our data structure.

Overall, each PLA-model forms a level of the PGM-index, and each model of that PLA-model forms a node of the data structure at that level, thus originating a sort of B-tree. The speciality of the PGM-index with respect to a B-tree is threefold: (i) the routing table at each node is given by a single linear model which guarantees constant space occupancy and logarithmic query time driven by the binary search over a subarray of size $2\varepsilon$; (ii) the fan-out of each node is variable and typically very large, depending on the "geometric complexity" of the set of indexed keys, so that the traversal of the PGM-index is fast; (iii) the PGM-index adapts its structure and its space occupancy to the distribution of the input keys, resulting as much independent as possible of their number. Figure 3 provides a pictorial example of a PGM-index built over $A$ and with a PLA-model of seven segment at the last level.

## 2.1 The piecewise linear model

Let us be given a sorted array $A = [k_1, k_2, \ldots, k_n]$ of $n$ (real and possibly repeated) keys drawn from a universe $\mathcal{U}$, and let us denote with $index : \mathcal{U} \to [1, n]$ the mapping from keys in $A$ to the position of their first occurrence in that array. We "extend" $index(k)$ to return the position of the *predecessor* of $k$ in $A$ if $k \notin A$.

The goal of this section is to describe an efficient implementation of $index(k)$ via a Piecewise Linear Approximation model, shortly referred hereafter with PLA-model. Given an integer $\varepsilon \geq 1$, we say that a PLA-model built on $A$ and with error $\varepsilon$ is a sequence of $m$ linear models (i.e. segments) $s_1, s_2, \ldots, s_m$, each one approximating up to error $\varepsilon$ the position of the keys in a subarray of $A$. Precisely, the $j$th segment $s_j$ is defined by a triple $(k_{i_j}, slope_j, intercept_j)$ that indexes the range of the universe keys $[k_{i_j}, k_{i_{j+1}})$ via a segment of slope $slope_j$ and intercept $intercept_j$. Each segment takes constant space to be stored, just two floats and one key, and it is used to approximate the position of any key $k \in [k_{i_j}, k_{i_{j+1}})$ by using the function $f_{s_j}(k) = k \times slope_j + intercept_j$, as depicted in Figure 2. We say that the position $f_{s_j}(k)$ is $\varepsilon$-approximate in the sense that the correct position $index(k)$ is no more than $\varepsilon$ positions away from the position $f_{s_j}(k)$ returned by the segment $s_j$. Therefore, each segment can be seen as an approximate predecessor search data structure for its covered range of keys (namely, $[k_{i_j}, k_{i_{j+1}})$) with constant query time and constant occupied space. We remark again that this is the interesting property of these segments because their space occupancy and access time do not depend on the size of the indexed range and thus, in turn, on the number of indexed keys.

Given the integer parameter $\varepsilon \geq 1$, the *piecewise linear $\varepsilon$-approximation problem* consists of computing the PLA-model which minimises the number of segments $m$, provided that each of them offers an $\varepsilon$-approximation of the positions of the keys in $A$ occurring in its covered range. The following Sections 2.1.1 and 2.1.2 describe a linear time and space algorithm that solves optimally this problem and provide some upper and lower bounds for $m$ and the number of points covered by each segment, respectively.

### 2.1.1 Optimal construction of the optimal PL-model

The piecewise linear approximation problem can be solved by dynamic programming in $O(n^3)$ time which is prohibitive on big data. Thus some authors [16] attacked this problem via a heuristic approach, called shrinking cone, which runs in linear time, but it does not guarantee any bound on the number of computed segments. Actually, these same authors showed examples of distributions of keys that are very worse for the efficacy of that algorithm.

It is interesting to notice that this problem has been extensively studied for lossy compression and similarity search of time series (see e.g. [32, 7, 10, 9, 42] and refs therein), and it admits streaming algorithms which take $O(n)$ optimal time. The key idea of this family of approaches is to reduce the $\varepsilon$-approximation problem to the one of constructing convex hulls of a set of points $S$, which in our case is $S = \{(k_i, index(k_i))\}_{i=1,\ldots,n}$. Then, the convex hull is enclosed in the thinnest rectangle (or strip) of height no more than $2\varepsilon$ from which the PLA-model is computed by taking the line which splits that rectangle in two halves.

For our application to the dictionary problem, we implemented the optimal algorithm of [42]. Theoretically, we will make then use of the following result:

▶ **Theorem 1.** *Let us be given a sequence $S = \{(x_i, y_i)\}_{i=1,\ldots,n}$ of two-dimensional points that are nondecreasing in their $x$-coordinates. There exists a streaming algorithm that in linear time and space computes the piecewise linear $\varepsilon$-approximation of $S$ formed by the minimum number of segments.*

### 2.1.2 Some simple bounds

In typical applications occurring mainly in the storage of posting lists of search engines or of the adjacency lists of graphs (e.g. Web graphs or Social Network graphs), the keys are $n$ increasing integers (i.e. the docIDs or the nodeIDs) which occur in a range $[d, d + U)$. We can give a tight upper bound on the minimum number of strips that are needed to cover the points as a function of $2\varepsilon$, the upper bound which we impose on their height.

When counting strips, the last one is a special case because it is not yet completed since it can be continued by adding other points to the right; to distinguish it from the other cases, we will call the other *full strips*. We use $\ell$ to denote the length of a full strip, from its beginning to the beginning of the next strip, and $m$ to denote the number of points it contains.

In Figure 4 we consider a worst-case situation in which the first $m$ points are aligned and consecutive in their $x$-coordinates, so that they impose the maximum slope in the final strip that covers them. Then we take the $(m + 1)$th point far away to the right at distance $\ell$ from the first point. We notice that the minimal strip including all these $m + 1$ points has the lower edge passing through the points $(0, 0)$ and $(\ell, m)$, and the upper edge parallel to the lower edge and passing through the point $(m - 1, m - 1)$. The height $h$ of that strip can be computed by simple geometric considerations, and it is $(m - 1)(\ell - m)/\ell$. By imposing that the current strip cannot include the $(m + 1)$ point and still guarantee a $2\varepsilon$-approximation, namely $h > 2\varepsilon$, we obtain $\ell > m(m - 1)/(m - 1 - 2\varepsilon)$.

Let us now study the function $f(m) = m(m - 1)/(m - 1 - 2\varepsilon)$ as our continuous lower bound to the length of a full strip with $m$ points. Our first observation is that the "average space" per point (namely, its average covered $x$-range) is at least $\frac{f(m)}{m} = 1 + 2\varepsilon/(m - 1 - 2\varepsilon)$, which is decreasing with $m$ as expected. The second observation is that $f$ has a minimum for $\hat{m} = \sqrt{2\varepsilon + 1}(\sqrt{2\varepsilon} + \sqrt{2\varepsilon + 1})$, and for that value it is $f(\hat{m}) = (\sqrt{2\varepsilon} + \sqrt{2\varepsilon + 1})^2$, so full strips will be longer than $8\varepsilon + 1$. But more importantly, we see that $f$ is a convex function, so in order to determine an upper bound to the number of full strips needed to cover all the points, we can consider all of them to have the same length and thus impose: $\lfloor n/m \rfloor f(m) + (n \bmod m) \leq U$. Focusing on the first addendum, which is the one of the full strips, we find $f(m) \leq (U\,m)/n$, and then substituting the definition above for $f(m)$, we derive $m \geq 1 + 2\varepsilon U/(U-n)$ which clearly holds *on average* over the maximum number of full strips created to cover $S$ with height at most $2\varepsilon$.

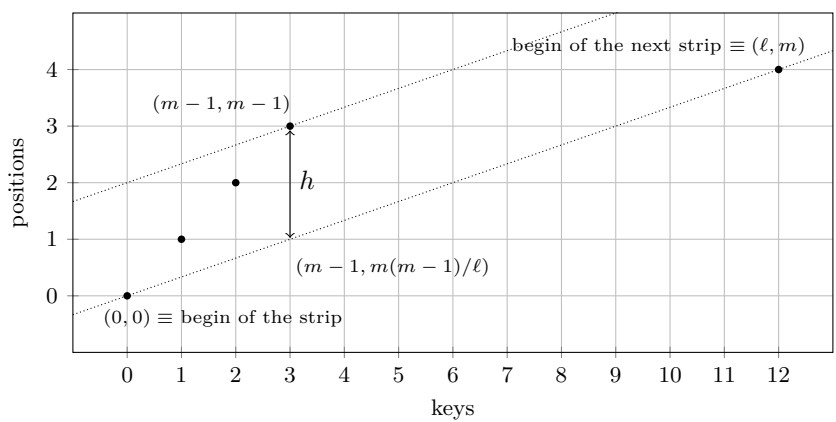

**Figure 4** Points inducing a full strip of height $h > 2\varepsilon$.

▶ **Lemma 2.** *Let us be given a sequence of $n$ two-dimensional points with integer coordinates, in a range of size $U$ such that their $x$-coordinates are distinct and sorted increasingly, and their $y$-coordinates grow of one unit at each point. The full strips will have length at least $8\varepsilon + 1$.*

*When the points are displaced in such a way to produce the maximum number of full strips, beholding the constraints on their range and their number, the optimal algorithm of Theorem 1 creates full strips which on average contain at least $1 + \frac{2\varepsilon U}{U-n}$ points each.*

We notice that as $U \to n$, all points have contiguous $x$-coordinate, and they are thus packed in a range of size $n$. Then, the lower bound diverges because we can fit all points of $S$ in one strip. If instead $U \to +\infty$, the lower bound converges to $1 + 2\varepsilon$. Actually, this lower bound holds for any $U$, even for points with repeated $x$-coordinate, so we have:

▶ **Corollary 3.** *Let us be given a sequence of $n$ two-dimensional points with integer coordinates, in a range of size $U$ such that their $x$-coordinates are sorted nondecreasingly, and their $y$-coordinates grow of one unit at each point. The optimal algorithm of Theorem 1 creates segments which contain at least $2\varepsilon$ points each.*

**Proof.** Any chunk of $2\varepsilon$ consecutive keys $k_i, k_{i+1}, \ldots, k_{i+2\varepsilon-1}$ in $A$ can be covered by the $\varepsilon$-approximate segment having null slope and intercept equal to $i + \varepsilon$ because those keys generate the points $(k_i, i), (k_{i+1}, i+1), \ldots, (k_{i+2\varepsilon-1}, i+2\varepsilon-1)$ and thus they have $y$-distance at most $\varepsilon$ from the line $y = i + \varepsilon$. ◀

From the previous calculations, it is also immediate to derive the desired upper bound on the number of strips.

▶ **Corollary 4.** *When the keys are $n$ distinct increasing integers occurring in a range $[d, d+U)$ with $n = \alpha U$, the optimal algorithm of Theorem 1 computes at most $\left\lceil \frac{n}{1 + \frac{2\varepsilon}{1-\alpha}} \right\rceil$ segments for an optimal piecewise linear $\varepsilon$-approximation of $S$.*

As a last comment we notice that, by allowing each key to be repeated at most $k$ times (which means vertical runs in $S$'s points when they are mapped to the Cartesian plane), the

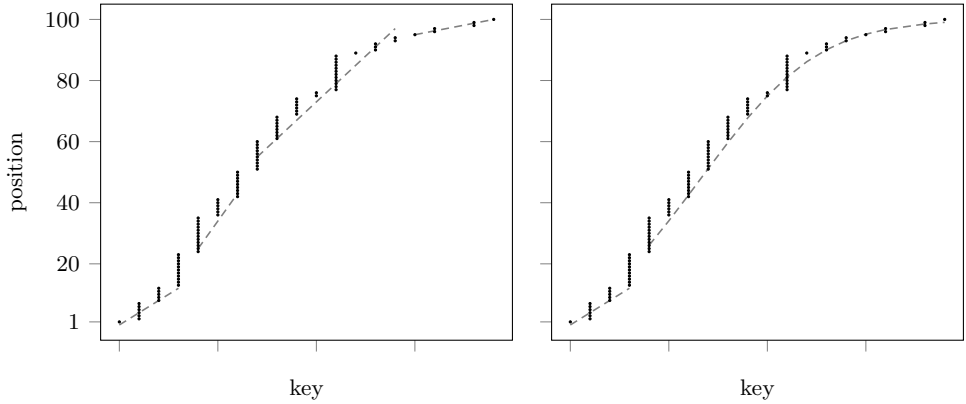

■ **Figure 5** The case of a PLA-model (left) vs a PNA-model (right): the former uses more "pieces" (i.e. 4 vs 2) to guarantee the same $\varepsilon$-approximation of the input sequence of 2D-points.

effect on the geometry is the possibility to stretch the coordinates of the points up to $k$ times, so that the upper bound in Corollary 4 above can be rewritten by substituting $\alpha$ with $\alpha/k$.

In Section 4.1, the lower bound of Corollary 3 will be used to model the space occupancy of a PGM-index, which is an essential ingredient of its multicriteria version.

## 2.2    Beyond linear models

In Section 2.1.1, we described an optimal solution to the piecewise linear $\varepsilon$-approximation problem, thus focusing on the $\varepsilon$-approximation of a sequence of points via segments. Nonetheless, it is reasonable to expect that more sophisticated models are more likely to cover larger regions of the key space due to their increased power in capturing complex data relationship. The net result could be a reduction in the number of models used in the piecewise approximation of the key space and hence a reduction of space overhead, as depicted in Figure 5. Another benefit could be that some kind of models, like neural networks, could take advantage of the speed-up given by GPUs, TPUs and other ML accelerators, which not only are ubiquitous (even in consumer hardware such as iPhone's Neural Engine or Google's Pixel Visual Core) but are also expected to be greatly improved in the future [23, 26].

For these reasons, we now turn our attention to the *general (i.e. nonlinear) piecewise $\varepsilon$-approximation problem* of a sequence of keys stored in an array $A$. In this problem, we are given an integer parameter $\varepsilon \geq 1$, a family of model types $\mathcal{M}$ sorted by complexity (i.e. space occupancy), and we are asked to compute the minimum number $m$ of models which $\varepsilon$-approximate the positions of the keys in $A$. The result is called Piecewise Nonlinear Approximation model (PNA-model), and it clearly generalises the PLA-model in that it uses both linear and nonlinear models in $\varepsilon$-approximating the positions of the input keys.

The approach we propose consists of refining in a *top-down fashion* the current PNA-model by adding more and more models until it guarantees the desired $\varepsilon$-approximation over all the keys in $A$. We call this algorithm TOP-DOWN-REGRESSION. Precisely, it starts building the PNA-model from the simplest model $f \in \mathcal{M}$ (i.e. the first one, and thus less complex model in the sorted order, i.e. the segment), trained on the whole array $A$. Then, it checks whether $f$ is an $\varepsilon$-approximation of $A$, i.e. it computes the absolute distance between the estimated position $f(k)$ and the true position $index(k)$ for each key $k \in A$. If every such distance is no more than $\varepsilon$, the algorithm stops and returns $f$. Otherwise, it picks the next (more complex) model in $\mathcal{M}$ and repeats the training. If no model in $\mathcal{M}$ is able to $\varepsilon$-approximate $A[1, n]$, then a breakpoint $p$ is suitably chosen, and the procedure is repeated recursively on the subarrays $A[1, p]$ and $A[p + 1, n]$.

The choice of the breakpoint $p$ for a subarray $A[a, b]$ is crucial for the efficiency and efficacy of the overall approach. Following [24], we could choose $p$ as the breakpoint that minimises the error made by the two models that cover $A[a, p]$ and $A[p + 1, b]$. However, the computation of $p$ would incur in a quadratic time cost because we should recompute the errors made by the two models for every breakpoint. This is prohibitive in time for the dataset sizes we will manage in our experiments.

Instead, we suggest to scan the array of errors made for the keys in $A[a, b]$ and place $p$ in a "problematic area" of $A$, intuitively, an area where the approximation error is too high or grows too much. In the experimental Section 5.1.2, we will explore different definitions of problematic area and find that the best choice is the one that puts two breakpoints at the start/end positions of the longest chain of errors larger than $\varepsilon$, i.e., consecutive keys of $A$ whose error in approximating their position with the simplest model is larger than $\varepsilon$.

At the end of the top-down phase, the algorithm scans the array of computed models and merges some neighbours if the approximation induced by the resulting merged model is no more than $\varepsilon$. The rationale of this last step is to limit the number of models created, as the placement of breakpoints could be too much greedy and thus overfit the distribution of the keys in $A$. For example, merging a deep neural network with its following segment $s$ consists of training the deep neural network also on the pairs $\{(k, index(k))\}_{k \in S}$, where $S$ is the set of keys covered by $s$. A model can be merged only with simpler models, not the other way around because if a region has been covered with a complex model, it means that it is too irregular to be covered by a simpler one.

The pseudocode of the algorithm is shown in Figure 6. Its actual time complexity depends on the input data. In the best case, the simplest (i.e. linear) model is enough to $\varepsilon$-approximate the whole array $A$, thus the time cost is $O(n)$. In the worst case the complexity is $O(n^3)$, but in practice it will result that the breakpoint taken at Line 14 of TOP-DOWN-REGRESSION will partition the subarrays into balanced parts thus inducing a time-complexity recurrence of $T(n) = 2T(n/2) + \Theta(n |\mathcal{M}|)$. Reasonably assuming that $|\mathcal{M}|$ is a constant, the recurrence has solution $T(n) = \Theta(n \log n)$. At the end, the procedure MERGE tries to merge the $m$ models $\varepsilon$-approximating array $A$, by calling procedure MODEL-MERGE and taking $O(nm)$ time. So we conclude that TOP-DOWN-REGRESSION runs in $O(n \log n + nm)$ time, under the assumptions that the subproblems have comparable size at every step and that $\mathcal{M}$ consists of a constant number of models.

## 2.3 Indexing piecewise (non)linear models

The two algorithms introduced in Sections 2.1.1 and 2.2 return a piecewise $\varepsilon$-approximation of the positions of the keys in the input array $A$ via a sequence $M$ of $m$ models that are either linear, as in the first algorithm, or possibly nonlinear, as in the second algorithm. Now, in order to solve the (static) dictionary problem, we need a way to find the model $f$ responsible for estimating the $\varepsilon$-approximate position $pos$ of a queried key $k$, namely, the rightmost model $f$ such that $f.key \leq k$. When $m$ is small, this search can be implemented by a linear scan over the attribute $f.key$. Otherwise, when $m$ is large, we need to binary search $M$ or index it via a proper data structure, such as a multi-way search tree such as B-tree or CSS-tree [34]. The membership query is thus answered in three steps. First, the multi-way search tree is queried to find the rightmost model $f$ such that $f.key \leq k$. Second, the model $f$ is used to *estimate* the position $pos = f(k)$ for the queried key $k$. Third, the exact position of $k$ is determined via a binary search within $A[pos - \varepsilon, pos + \varepsilon]$. It is possible

Top-Down-Regression$(A, \varepsilon, \mathcal{M})$

  1  let *models* be an empty list
  2  let *stack* be an empty stack
  3  Push$(stack, (1, n))$
  4  **while** not Empty$(stack)$
  5    $a, b =$ Pop$(stack)$
  6    **for** each Model $\in \mathcal{M}$
  7      $f =$ **new** Model$(A, a, b)$
  8      $errors =$ Compute-Errors$(A, a, b, f)$
  9      **if** Max$(errors) \leq \varepsilon$
 10        **break**
 11    **if** Max$(errors) \leq \varepsilon$
 12      List-Append$(models, f)$
 13    **else**
 14      choose a breakpoint $p$ for $A[a, b]$
 15      Push$(stack, (p + 1, b))$
 16      Push$(stack, (a, p))$
 17  **return** Merge$(A, \varepsilon, models)$

Merge$(A, \varepsilon, models)$

  1  let *merged* be a list containing only $models[1]$
  2  **for** $i = 2$ **to** Size$(models)$
  3    let *tail* be the last item in *merged*
  4    **if** not *tail*.Is-Mergeable$(models[i])$
  5      List-Append$(merged, models[i])$
  6    **else**
  7      $f = tail$.Model-Merge$(models[i])$
  8      **if** Compute-Errors$(models[i]) \leq \varepsilon$
  9        List-Delete-Tail$(merged)$
 10        List-Append$(merged, f)$
 11      **else**
 12        List-Append$(merged, models[i])$
 13  **return** *merged*

🟨 **Figure 6** The algorithm for the piecewise nonlinear $\varepsilon$-approximation problem. The procedure Model$(A, a, b)$ trains a model $f \in \mathcal{M}$ on $A[a, b]$, while the procedure Compute-Errors$(A, a, b, f)$ returns an array of size $b - a + 1$ whose $i$th item is the error induced by model $f$ in estimating the position of key $k = A[a + i - 1]$, namely, $|\lfloor f(k) \rfloor - index(k)|$.

to support also successor, predecessor[3] and range queries with simple adjustments to the final step.

The combination of an indexing data structure with the piecewise (linear or nonlinear) $\varepsilon$-approximation of the positions of the keys in $A$ constitutes our novel index, called PGM-index, which stands for *Piecewise Geometric Model* index (see Figure 3 for a pictorial example).

▶ **Theorem 5.** *The PGM-index takes $\Theta(m)$ space and answers membership, successor and predecessor queries in $O(t_{\text{index}}(m) + \log \varepsilon)$ time, where the integer $\varepsilon \geq 1$ denotes the error guaranteed in the approximation of positions of keys in $A[1, n]$ by the $m$ (linear or nonlinear) models created over them, and $t_{\text{index}}(m)$ is the time needed to find the model responsible for the queried key by using the data structure built over those $m$ models.*

The efficiency in time and space of the PGM-index strictly relies on the value of $m$ which is smaller than or equal to $n/(2\varepsilon)$ because of Corollary 3. Hence, $m < n$ given that $\varepsilon \geq 1$. In practice, as we will show in the experimental Section 5.1, $m$ is significantly smaller than $n$ so that the space savings are up to several orders of magnitude. This will impact onto $t_{\text{index}}(m)$ too, and thus onto the efficiency of the query operations.

However, the indexing strategy above does not take full advantage of the Piecewise Geometric Modelling of a sequence of keys that we are introducing in this paper because it resorts a classic data structure to index $M$. Therefore, we introduce a third (successful) strategy to index $M$ which consists of repeating the Piecewise Geometric Modelling process *recursively* on a proper set of keys derived from the previous modelling steps until there is only one model left. More precisely, we start with the sequence of models $M$ constructed over the whole input array $A$, we then take the first key of $A$ covered by each model in $M$ and finally construct another (and smaller) PNA-model over those $m = |M|$ keys. We proceed this way recursively until the PNA-model consists of one unique model. Each PNA-model

---

[3] Defined as $successor(q) = k_i$ such that $k_{i-1} < q \leq k_i$. Symmetrically, $predecessor(q) = k_i$ such that $k_i \leq q < k_{i+1}$.

RECURSIVE-GEOMETRIC-MODELLING($A, \varepsilon$)

1   let *levels* be an empty list
2   *keys* = $A$
3   **repeat**
4       $l$ = BUILD-PIECEWISE-MODEL(*keys*, $\varepsilon$)
5       LIST-PREPEND(*levels*, $l$)
6       $m$ = SIZE($l$)
7       *keys* = $[\, l[1].key, \ldots, l[m].key \,]$
8   **until** $m = 1$
9   **return** *levels*

QUERY-R($A, \varepsilon, levels, k$)

1   $pos$ = *levels*[1][1]($k$)
2   **for** $i = 2$ **to** SIZE(*levels*)
3       $lo$ = max$\{pos - \varepsilon, 1\}$
4       $hi$ = min$\{pos + \varepsilon, $ SIZE(*levels*[$i$])$\}$
5       $f$ = rightmost model $f'$ in *levels*[$i$][$lo, hi$]
            s.t. $f'.key \leq k$, found by binary search
6       $g$ = the model at the right of $f$
7       $pos$ = min$\{\lfloor f(k) \rfloor, \lfloor g(g.key) \rfloor\}$
8   $lo$ = max$\{pos - \varepsilon, 1\}$
9   $hi$ = min$\{pos + \varepsilon, n\}$
10  **return** binary search for $k$ in $A[lo, hi]$

**Figure 7** The pseudocode to build and query a PGM-index where the models are in turn indexed by several recursive levels. BUILD-PIECEWISE-MODEL is either the algorithm described in Section 2.1.1 to build linear models, or the one in Section 2.2 to build (non)linear models that $\varepsilon$-approximate the positions of the keys in $A$.

will form a level of the PGM-index, and each model of that PNA-model will form a node of the data structure at that level, thus forming a sort of B-tree. The speciality of this *Recursive PGM-index* is threefold: (i) the routing table at each B-tree node is given by a single model which guarantees, unlike classic arrays in B-trees, constant space occupancy and logarithmic query time driven by the parameter $\varepsilon$; (ii) the fan-out of each node, unlike classic B-trees, is variable and typically very large, depending on the $\varepsilon$-approximability of the positions of the indexed keys; (iii) given the previous point (ii), the depth of the B-tree will be usually very small, and in general the Recursive PGM-index will adapt its structure and its space occupancy to the "geometric distribution" of the input keys resulting highly compressed, as we will show in the experimental section.

The membership query over the Recursive PGM-index works as follows. At every level, we take the model referring to the visited node and use it to estimate the position of the searched key $k$ among the keys of the next level. The real position is then found by a binary search in a range of size $2\varepsilon$ centred around the estimated position. Given that every key on the next level is the first key covered by a node on that level, we have identified the next node to visit, and the process continues until the last level of $A$'s keys is reached.[4] The pseudocode of this construction process, which we call RECURSIVE-GEOMETRIC-MODELLING, is shown in Figure 7 together with the procedure QUERY-R that implements the search for a key into the obtained Recursive PGM-index. For a pictorial example of a Recursive PGM-index, we refer the reader to the previous Figure 3.

We observe that, when working in hierarchical memories, it could be advantageous to modify RECURSIVE-GEOMETRIC-MODELLING to use a different $\varepsilon_\ell$ in each level $\ell \in [1, L]$ of the Recursive PGM-index. In fact, we could fit the first $(L-1)$ levels in the cache and set $\varepsilon_\ell$ to the cache-line size, and we could store the input array $A$ in the secondary memory and set $\varepsilon_L$ to a multiple of the disk-page size.

▶ **Theorem 6.** *The Recursive PGM-index takes $\Theta(m)$ space and answers membership, successor and predecessor queries in $O(\log m)$ time. In the External Memory (EM) model,* queries take $O((\log_c m) \log(\varepsilon/B))$ I/Os, where the integer $\varepsilon \geq 1$ denotes the error guaranteed in the approximation of the positions of keys in $A[1, n]$ by the m (linear or nonlinear) models created over them, and $c \geq 2\varepsilon$ denotes the fan-out of the data structure.

---

[4] Some care has to be taken to correctly approximate the positions of keys falling between the last key covered by a model $f$ and the first key (i.e. $g.key$) covered by the model $g$ at the right of $f$. This is the case of keys that do not occur in $A$, and for which we are asking to search for their predecessor or successor. The subtle issue is that, for these keys, model $f$ is not guaranteed to return an $\varepsilon$-approximate position in $A$. This subtlety is managed by Line 7 of procedure QUERY-R (shown in Figure 7).

**Proof.** Each step of Recursive-Geometric-Modelling reduces the number of models by a variable factor $c$ which is nonetheless larger than $2\varepsilon$ because of Corollary 3. The number of levels is, therefore, $L = O(\log_c m)$, and the total space required by the recursive geometric model is $\sum_{\ell=0}^{L} m/\varepsilon^\ell = \Theta(m)$. For the membership query, the bounds on the running time and the I/O complexity follow easily by observing that Query-R performs $L$ binary searches over intervals having size at most $2\varepsilon$. ◄

Theorem 6 highlights the main novelty of the PGM-index: its space overhead does not grow linearly with $n$, as in the traditional indexes mentioned in Section 1, but it grows with the "geometric complexity" of the input array $A$ and decreases with the value of $\varepsilon$. From a theoretical perspective, the PNA-model at the last level of a PGM-index cannot have more models than $n/(2\varepsilon)$, hence $m < n$ given that $\varepsilon \geq 1$ (see Corollary 3). And since this fact holds for the recursive levels too, it follows that asymptotically the PGM-index cannot be worse in space and time than a $2\varepsilon$-way tree (just take $c = 2\varepsilon = \Theta(B)$ in Theorem 6). In practice, as we will show in the experimental Section 5.2, the PGM-index is much faster and succinct than a $2\varepsilon$-way tree due to the use of segments as routing tables, and due to the large variable fan-out $c$ they allow to achieve. For instance, a query takes $O(\log(\varepsilon/B)) = O(1)$ I/Os because typically in practice $c \gg 2\varepsilon$.

Other than traditional tree-based indexes, the PGM-index generalises (i) the A-tree [16], in that it recursively and level-wise applies linear models to form a tree structure of (possibly) large fan-out, and (ii) the RMI [27], in that the tree structure of the PGM-index is not fixed in advance, but adapts itself to the distribution of the keys available at every level. Consequently, the number of levels, the number of nodes at every level, and the fan-out of each node are not fixed but they are learned from the input data as a function of the approximate error $\varepsilon$. The Multicriteria framework, explained in Section 4, will show how to control in a principled way the parameter $\varepsilon$, and thus the space and the time efficiency of the PGM-index.

Table 1 summarises the query costs, both in the RAM model and in the EM model (with page size $B$), of the three strategies described in this section. In the case of range queries, time costs increase by the output-sensitive term of $O(K)$ (in EM model, $O(K/B)$), where $K$ is the number of keys satisfying the range query. To appreciate the computational efficiency of the Recursive PGM-index, we notice that in theory $L = O(\log_\varepsilon m)$ and $m = O(n/\varepsilon)$ with $\varepsilon \geq 1$, but in practice those upper bounds are very pessimistic in that $L$ turns out to be a very small constant and $m$ results many orders of magnitude smaller than $n$. The thorough experimental results of Section 5 performed on large datasets will support these theoretical considerations by showing improvements of several orders of magnitude in space and significant improvements in time. This is the reason why we prefer to leave the bounds as a function of $m$ rather than $n$.[5]

---

[5] As a practical consideration, the PGM-index indexed the biggest and most complex dataset available to us of about 715M items with $m = 423$K linear models and taking $\varepsilon = 64$ (see Section 5.1.1). This is a reduction in the index size of three orders of magnitude. We also found in our experiments that $m$ decreases as a power of $\varepsilon$, thus making the upper levels of the Recursive PGM-index pretty much negligible in terms of occupied space.

| Data structure | Space | RAM model worst case time | EM model worst case I/Os | EM model best case I/Os |
|---|---|---|---|---|
| Plain sorted array | $O(1)$ | $O(\log n)$ | $O(\log \frac{n}{B})$ | $O(\log \frac{n}{B})$ |
| Multiway tree (e.g., B-tree) | $\Theta(n)$ | $O(\log_B n)$ | $O(\log_B n)$ | $O(\log_B n)$ |
| PGM-index w. binary search | $\Theta(m)$ | $O(\log m + \log \varepsilon)$ | $O(\log \frac{m}{B})$ | $O(\log \frac{m}{B})$ |
| PGM-index w. multiway tree | $\Theta(m)$ | $O(\log m + \log \varepsilon)$ | $O(\log_B m)$ | $O(\log_B m)$ |
| Recursive PGM-index | $\Theta(m)$ | $O(\log m)$ | $O(\log_c m)$ $c \geq 2\varepsilon = \Omega(B)$ | $O(1)$ |

🟨 **Table 1** Time and I/O complexity of the predecessor query operation with traditional data structures, and the three variants of PGM-index described in Section 2.3. Recall that the integer $\varepsilon \geq 1$ denotes the error guaranteed in the approximation of the positions of keys in $A$ by the $m$ (linear or nonlinear) models. In the EM model we assumed $\varepsilon = \Theta(B)$. In theory, $m \leq n/(2\varepsilon)$, but in practice this upper bound is very pessimistic in that $m$ is many orders of magnitude smaller than $n$ (see Section 5).

## 🟧 3 The Distribution-Aware PGM-index

The PGM-index of Theorem 6 implicitly assumes that queries are uniformly distributed, but this seldom happens in practice. For example, in search engines queries are very well known to follow very skewed distributions such as the Zipf's law [41]. In such cases, it is desirable to have an index that answers the most frequent queries faster than the rare ones, so to achieve a higher query throughput. Previous work exploited query distribution in the design of binary trees [5, 25], treaps [36], skip lists [3], and text indexes [13], just to cite a few.

In this section, we introduce an orthogonal and very simple approach that builds upon the PGM-index by proposing a variant that adapts itself not only to the distribution of the keys but also to the distribution of the queries. This turns out to be the *first* distribution-aware learned index to date, with the additional positive feature of being very succinct in space.

Formally speaking, given a sequence of weighted keys $S = \{(k_1, p_1), \ldots, (k_n, p_n)\}$, where $p_i$ is the probability to query the key $k_i$, we want to solve the distribution-aware dictionary problem, which asks for a data structure that searches for a key $k_i$ in time $O(\log(1/p_i))$, so that the average query time coincides with the entropy of the query distribution $\mathcal{H} = \sum_{i=1,\ldots,n} p_i \log(1/p_i)$.

We recall that O'Rourke [32] proposed an algorithm that, given a $y$-range for each one of $n$ points in the plane, finds the set of all directions that intersect those ranges in $O(n)$ time. This algorithm can be turned into an optimal one which finds the minimum number of segments intersecting the $y$-range of those keys, via an incremental approach that starts a new direction (i.e. segment) as soon as the set of possible directions intersecting a given subset of points is empty [42].

Our key idea is then to define, for every key $k_i$, a $y$-range of size $y_i = \min\{1/p_i, \varepsilon\}$, for a given integer parameter $\varepsilon \geq 1$, and then apply O'Rourke's algorithm on that set of keys and ranges. Clearly, for the keys whose $y$-range is $\varepsilon$ we can use Theorem 6 and derive the same space bound of $O(m)$; whereas for the keys whose $y$-range is $1/p_i < \varepsilon$ we observe that these keys are no more than $\varepsilon$ and the $p_i$s sum up to 1, so they induce in the worst case $2\varepsilon$ extra segments. Therefore, the total space occupancy of the leaf level of the recursive distribution-aware index is $\Theta(m + \varepsilon)$, where $m$ is the one defined in Theorem 6. Now, let us assume that the search for a key $k_i$ arrived at the last level of this PGM-index and thus we know in which segment to search for $k_i$: the final binary search step within the approximate range returned by that segment takes $O(\log \min\{1/p_i, \varepsilon\}) = O(\log(1/p_i))$ as we aimed for.[6]

But how do we find that segment in a distribution-aware manner? We proceed as in the recursive PGM-index but, here, we need to be very careful in designing the recursive step because of the probabilities (and thus the variable $y$-ranges) assigned to the recursively defined set of keys.

Let us consider the segment covering the range of keys $S_{[a,b]} = \{(k_a, p_a), \ldots, (k_b, p_b)\}$, and denote by $q_{a,b} = \max_{i \in [a,b]} p_i$ the maximum probability of a key in $S_{[a,b]}$, and by $P_{a,b} = \sum_{i=a}^{b} p_i$ the cumulative probability of all keys in $S_{[a,b]}$ (which is indeed the probability to end up in that segment when searching for one of its keys). We create the new set $S' = \{\ldots, (k_a, q_{a,b}/P_{a,b}), \ldots\}$ formed by the first key $k_a$ covered by each segment (as in the recursive PGM-index) and setting its associated probability to $q_{a,b}/P_{a,b}$. Then, we construct the next upper level of the PGM-index by applying O'Rourke's algorithm on $S'$. If we iterate the above analysis for this new level of weighted segments, we conclude that: if we know from the search executed on the levels above that $k_i \in S_{[a,b]}$, the time cost to search for $k_i$ in this level is $O(\log \min\{P_{a,b}/q_{a,b}, \varepsilon\}) = O(\log(P_{a,b}/p_i))$.

Let us repeat this argument for another upper level in order to understand the "structure" of the search time complexity. Let us denote the segment that covers the range of keys which include $k_i$ with $S_{[a',b']} \supseteq S_{[a,b]}$, the cumulative probability with $P_{a',b'}$, and thus assign to its first key $k_{a'}$ the probability $r/P_{a',b'}$, where $r$ is the maximum probability of the form $P_{a,b}$ of the ranges included in $[a',b']$; in other words, if $[a',b']$ is partitioned into $[z_1, \ldots, z_c]$, then $r = \max_{i \in [1,c)} P_{z_i, z_{i+1}}$. Reasoning as done previously, if we know from the search executed on the levels above that $k_i \in S_{[a',b']}$, the time cost to search for $k_i$ in the this level is $O(\log \min\{P_{a',b'}/r, \varepsilon\}) = O(\log(P_{a',b'}/P_{a,b}))$ because $[a,b]$ is, by definition, one of these ranges in which $[a',b']$ is partitioned.

Repeating this design until one single segment is obtained (whose cumulative probability is one), we get a total time cost for the search in all levels of the PGM-index but the last one which is equal to a sum of logarithms whose arguments "cancel out" and get $O(\log(1/p_i))$.

As far as the space bound is concerned, we recall that the number of levels in the PGM-index is $L = O(\log_\varepsilon m)$ and that we have to account for the $\varepsilon$-extra segments per level returned by the O'Rourke's algorithm. Consequently, this distribution-aware variant of the PGM-index takes $O(m + L\varepsilon) = O(m + \varepsilon \log_\varepsilon m)$ space. This space bound is indeed $O(m)$ because $\varepsilon$ is a constant parameter with typically very small values in practice (see Section 5.1). We have thus proved the following:[7]

▶ **Theorem 7.** *The Distribution-Aware PGM-index takes $O(m)$ space and solves the distribution-aware dictionary problem in $O(\mathcal{H})$ average time, where $\mathcal{H}$ is entropy of the query distribution and $m$ is the optimally minimum number of segments created over the $n$ input keys (e.g. by O'Rourke's algorithm [32]).*

As we repeatedly observed in the previous sections (see also Table 1), in theory $m \leq n/(2\varepsilon)$, but in practice this upper bound is very pessimistic because $m$ is many orders of magnitude smaller than $n$. This means that our distribution-aware PGM-index offers the very precious feature of being very succinct in its space occupancy.

---

[6] In an actual implementation, one would avoid storing the $y_i$s that delimit the binary search ranges and perform instead an exponential search.

[7] We recall that, with respect to hashing, our solution can be extended to answer predecessor queries on keys of the universe not present in the dictionary; moreover, it takes an extra space, other than the keys, which depends on $m \ll n$ and thus turns out to be very succinct in practice.

## <span style="background:yellow">4</span>    The multicriteria version of the PGM-index

Tuning a data structure to match the application's needs is often a difficult and error-prone task for a software engineer, not to mention that these needs may change over time due to mutations in data distribution, devices, resource requirements, and so on. The typical approach is to grid search the various instances of the data structure to be tuned until the one that matches the application's needs is found. As an example, the RMI of [27] has two dimensions to tune (the number of stages, i.e. levels of its DAG structure, and the number of models in each stage), with a possibly huge search space to explore. Hence the software engineer has to restrict the large search space of the parameters in order to make the grid search efficient, and this may potentially miss the optimal solution.

In the following two subsections we show that this tuning process can be efficiently automated over the PGM-index via an optimisation strategy that: (i) given a space constraint outputs the PGM-index that minimises its query time (Section 4.1); symmetrically, (ii) given a maximum query time outputs the PGM-index that minimises its occupied space (Section 4.2).

### 4.1    The time-minimisation problem

We start by modelling the query time and the space occupancy of the Recursive PGM-index as functions of the $\varepsilon$ parameter. The former can be modelled, according to Theorem 6, as $t(\varepsilon) = c(\log_{2\varepsilon} m) \log(2\varepsilon/B)$, where $B$ is the page size of the EM model, $m$ is the number of models in the last level of the PGM-index, and $c$ depends on the access latency of the memory, which we assume constant for simplicity. For the latter, we introduce $s_\ell(\varepsilon)$, which denotes the number of models needed to have accuracy $\varepsilon$ over the positions of the keys available at level $\ell$ of the Recursive PGM-index, and we compute the overall number of models as $s(\varepsilon) = \sum_{\ell=1}^{L} s_\ell(\varepsilon)$. By Corollary 3, we know that $s_L(\varepsilon) = m \leq n/(2\varepsilon)$ for any $\varepsilon \geq 1$ and that $s_{\ell-1}(\varepsilon) \leq s_\ell(\varepsilon)/(2\varepsilon)$. So that $s(\varepsilon) \leq \sum_{\ell=0}^{L} m/(2\varepsilon)^\ell = (2\varepsilon m - 1)/(2\varepsilon - 1)$.

Given a space bound $s_{max}$, the problem is to minimise $t(\varepsilon)$ subject to $s(\varepsilon) \leq s_{max}$.[8] The greatest challenge of this problem is that we do not have a closed formula for $s(\varepsilon)$, but only an upper bound which does not depend on the underlying dataset as instead $s(\varepsilon)$ does. Section 5.1.1 will show that in practice we can choose to model $m = s_L(\varepsilon)$ with a simple power-law relation having the general form $a\varepsilon^{-b}$, whose parameters $a$ and $b$ will be properly estimated on the dataset at hand. This modelling covers both the rather pessimistic case described by Corollary 3 (just take $a = n/2$ and $b = 1$) and the lucky one in which the dataset is strictly linear (just take $a$ small and $b$ equal to zero).

**Solving the time-minimisation problem.** The time-minimisation problem is illustrated in Figure 8. As one would expect, the query time $t(\varepsilon)$ of the Recursive PGM-index increases with increasing $\varepsilon$, since the area $[pos - \varepsilon, pos + \varepsilon]$ gets larger. Higher values of $s_{max}$, instead, push to the left the vertical dashed line, which delimits the feasible area. As a matter of fact, with more space we are able to introduce more models and thus have a more accurate index. It should be clear from the figure that solving the time-minimisation problem reduces to finding the position of the vertical dashed line, i.e., the value of $\varepsilon$, for which $s(\varepsilon) = s_{max}$ because it is the lowest $\varepsilon$ that we can afford. Since $s(\varepsilon)$ is monotonically decreasing, such value of $\varepsilon$ can be found by a binary search in the bounded interval $\mathcal{E} = [B/2, n/2]$ which is derived by requiring that each model errs at least a page size (i.e. $2\varepsilon \geq B$), since lower $\varepsilon$ values do not save I/Os, and by observing that one model is the minimum possible space (i.e. $2\varepsilon \leq n$, according to Corollary 3).

---

[8] Throughout this section, we assume that a disk page contains exactly $B$ keys. This assumption, which simplifies our formulas, can be relaxed by putting the proper machine- and application-dependent constants in front of $t(\varepsilon)$ and $s(\varepsilon)$. For an actual implementation it would be more appropriate to express $s(\varepsilon)$ and $s_{max}$ in bytes, as we will do in the experiments described in Section 5.3.

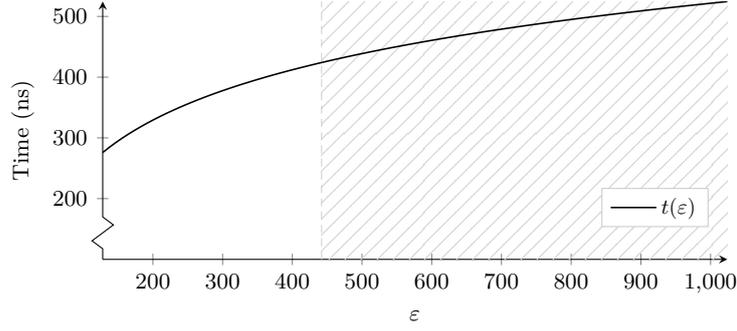

**Figure 8** A graphical description of the time-minimisation problem. The solid line represents the function $t(\varepsilon)$, while the area highlighted by diagonal lines represents feasible values for $\varepsilon$, i.e., the ones for which $s(\varepsilon) \leq s_{max}$.

**Speeding up the search.** Provided that our power-law approximation holds, we show that it is possible to speed up the solution based on binary search by suitably guessing the next value of $\varepsilon$ rather than taking the midpoint of the current search interval. In fact, we can find the root of $s(\varepsilon) - s_{max}$, i.e. the value $\varepsilon_g$ for which $s(\varepsilon_g) = s_{max}$. We emphasise that such $\varepsilon_g$ may not be the solution of our problem, as it may be the case that the approximation or the fitting of $s(\varepsilon)$ by means of a power-law is not precise. Thus, more iterations of the search may be needed to find the optimum $\varepsilon$ value. Our experiments will show that actually the number of guesses is sufficiently small that the approach will be very fast in practice, surely faster than binary search.

Since the method of guessing the next $\varepsilon$ closely resembles interpolation search, we adopt the method described by [18] in order to further limit useless steps: we gradually switch to a standard binary search by biasing the guess $\varepsilon_g$ towards the midpoint of the current search range. To be specific, let $|\mathcal{E}|$ be the size of our search space. Let *guesses* be the number of times the guess $\varepsilon_g$ was used to choose the next iterate during the execution of the algorithm, and *threshold* $= \lceil \log \log |\mathcal{E}| \rceil$ be the maximum number of guesses allowed. Suppose that $\varepsilon_m$ and $\varepsilon_g$ are, respectively, the current midpoint of the binary search and the guess. If *guesses* < *threshold*, we choose the next iterate of $\varepsilon$ as $\varrho \times \varepsilon_m + (1 - \varrho) \times \varepsilon_g$, where $\varrho = guesses/threshold$. Otherwise, we choose the next iterate of $\varepsilon$ as $\varepsilon_m$. This gradual transition from $\varepsilon_g$ to $\varepsilon_m$ avoids the deterioration to linear search, which occurs when the true $s(\varepsilon)$ is different from the approximation that we use as the guess and which, we must remark, was observed empirically in our diverse albeit small collection of datasets. As a result, with such a choice of the next iterate, we have the same worst-case performance of a binary search even when the guess $\varepsilon_g$ is wrong.

To summarise, we search the value $\varepsilon$ giving the smallest query time $t(\varepsilon)$ within space bound $s_{max}$ as follows:

1. We run several iterations of binary search (e.g., 4 or 5), and save in a set $\mathcal{D}$ the pairs $(\varepsilon_i, s_L(\varepsilon_i))$ obtained at each step by running BUILD-PIECEWISE-MODEL.
2. We fit a power-law of the form $a\varepsilon^{-b}$ on the set of points $\mathcal{D}$, thus determining the values of the parameters $a$ and $b$. Then we compute a goodness-of-fit measure, such as $R^2$.

**3.** If the fitted model is good enough (i.e., $R^2$ is close to one), we choose the next iterate of $\varepsilon$ using the biased guess described in the text above.

**4.** Otherwise, we jump to 1 and continue with the standard binary search, with the hope that collecting new pairs in $\mathcal{D}$ improves the fitting of the power-law.

Note that the cost of the fitting at Step 2 is negligible with respect to the index construction at Step 1, since it is CPU bounded and not I/O bounded. In practice, the number of iterations and index constructions that Step 3 saves with respect to a plain binary search is significant, as we will see in the experimental Section 5.3.

## 4.2    The space-minimisation problem

Given a time bound $t_{max}$, the space-minimisation problem consists of minimising $s(\varepsilon)$ subject to $t(\varepsilon) \leq t_{max}$. As for the problem in the previous section, we can binary search inside the interval $\mathcal{E}$ and look for the maximum $\varepsilon$ which satisfies the constraint. Likewise, we could guess the next iterate for $\varepsilon$ by solving the equation $t(\varepsilon) = t_{max}$, that is solving $c(\log_{2\varepsilon} s_L(\varepsilon)) \log(2\varepsilon/B) = t_{max}$, in which $s_L(\varepsilon)$ is replaced by the power-law approximation $a\varepsilon^{-b}$ for proper $a$ and $b$, and $c$ is replaced by the measured memory latency of the given machine. However, this approach raises a subtle issue: the time model could not be a correct estimate of the true (empirical) query time because of hardware-dependent factors such as the presence of several caches and the CPU pre-fetching. To further complicate this issue, we note that both $s(\varepsilon)$ and $t(\varepsilon)$ depend on the power-law approximation $a\varepsilon^{-b}$.

For these reasons, instead of using the time model $t(\varepsilon)$ to steer the search, we suggest to measure and use the actual average query time $\bar{t}(\varepsilon)$ of the PGM-index over a fixed batch of random queries. Moreover, instead of binary searching inside the whole $\mathcal{E}$, we run an exponential search starting from the solution of the dominating term $c\log(2\varepsilon/B) = t_{max}$, i.e. the cost of searching the data. Eventually, we stop the search of the best $\varepsilon$ as soon as the searched range is smaller than a given threshold because $\bar{t}(\varepsilon)$ is subject to measurement errors (e.g. due to an unpredictable CPU scheduler).

## 5    Experiments

We experimented with an implementation in C++ of the algorithms described in this paper on a machine with a 2.3 GHz Intel Xeon Gold and 192 GiB memory.[9] We used the following three datasets, each having different data distributions, regularities and patterns:

**1.** *Web logs* [27]. Contains timestamps of about 715M requests to a web server of a computer science department. This dataset exhibits several patterns, for example, fewer requests during summer and night time.

**2.** *Longitude* [31]. Contains longitudes of about 166M points-of-interest from OpenStreetMap.

**3.** *IoT* [27]. Contains timestamps of about 26M events recorded by IoT sensors (e.g., door, motion, power) installed throughout an academic building. Trends in this dataset generally reflect human activity, such as a high concentration of events during daytime and semesters.

---

[9] The implementation will be released soon.

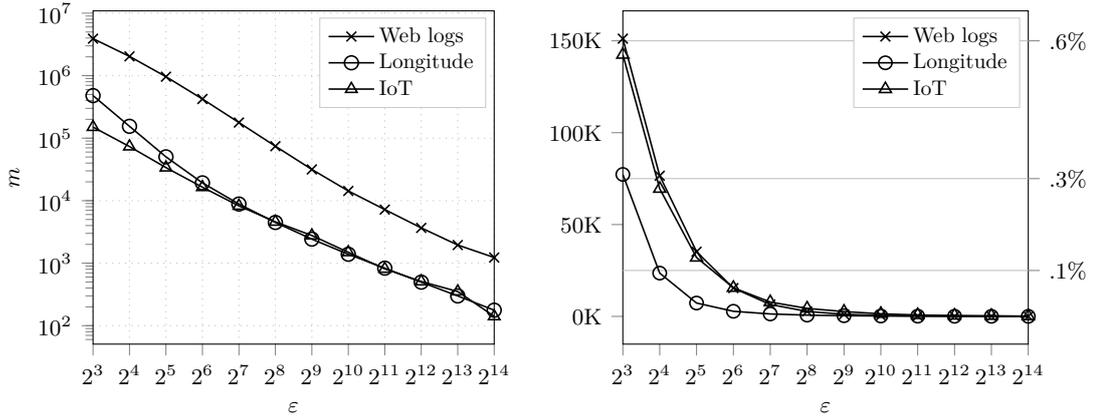

**(a)** $m$ over the whole datasets.

**(b)** $m$ over the first 25M entries of each dataset.

**Figure 9** Number of segments $m$ in the optimal PLA-model for several values of $\varepsilon$. The ratio $m/n$ is shown in percentage on the right of the right chart.

## 5.1   Piecewise geometric model

In this batch of experiments, we evaluated the goodness of the PNA-model in terms of the number of models it uses to $\varepsilon$-approximate the keys in the three datasets above. Savings in the number of models impact onto the space occupancy but also onto the query time of the PGM-index, as there are fewer comparisons to find the segment responsible for a key. Section 5.1.1 focuses on the PLA-model, while Section 5.1.2 focuses on the PNA-model with both linear and nonlinear models.

### 5.1.1   Piecewise linear model

Figure 9a shows a log-log plot of the size $m$ of the PLA-model built by the optimal construction algorithm of [42] for several given values of $\varepsilon$ (see Section 2.1.1). Even when $\varepsilon$ is as little as 8, the number $m$ of segments is more than two orders of magnitude smaller than the original datasets size $n$. To dig into the "geometric complexity" of the three datasets, we repeated the previous experiment on the first 25M entries and discovered that Web logs and IoT are the most complex to index because they require more segments than Longitude (Figure 9b).

In order to understand the impact of the optimal construction algorithm of [42] onto the space occupancy of the PGM-index, we compared its number of computed segments against the currently best and heuristic algorithm proposed in [16, 37], called shrinking cone (shortly, SC). The optimal algorithm significantly improves SC by reducing the number of segments of about 38%–63.3% (Figure 10) while maintaining the same time efficiency in practice and optimality in asymptotic time complexity.

Then, for designing a multicriteria PGM-index (see Section 4 and Section 5.3), we studied the behaviour of $m = s_L(\varepsilon)$ by varying $\varepsilon$. We fitted ninety different functions over about two hundred points $(\varepsilon, s_L(\varepsilon))$ generated beforehand by a long-running grid search over our datasets. At the end, looking at the fittings, we chose to model $s_L(\varepsilon)$ with a simple *power law* having the general form $a\varepsilon^{-b}$.

For completeness, we report that the algorithm with $\varepsilon = 8$ built a PLA-model for Web logs in 2.59 seconds, whereas it took less than 1 second for Longitude and IoT datasets.

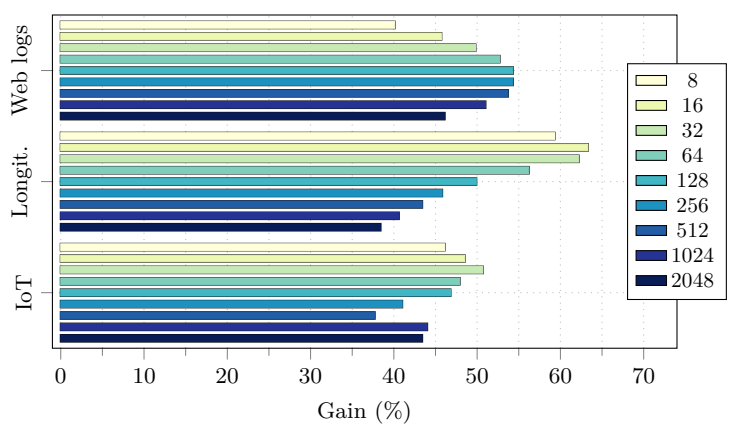

■ **Figure 10** Gain in number of segments with respect to SC [16] for various values of $\varepsilon$.

▶ **Summary.**  *The optimal algorithm [42] to construct the PLA-model is very fast (i.e. less than 3 seconds for the about 715M keys of the Web logs dataset) and induces a significant gain in the space occupancy of the PGM-index, which indeed results up to 60% more succinct than what was achieved by previously known heuristic algorithms (i.e. SC [16, 37]), and overall from two to five orders of magnitude smaller than the original datasets size.*

### 5.1.2    Piecewise nonlinear model

We experimented with an implementation of the TOP-DOWN-REGRESSION algorithm described in Section 2.2. The pool of models consisted either of segments or of three neural networks selected with a random search of neural networks with 0 hidden layers or 1 hidden layer with 2–5 hidden units, and activations in $\{elu, softplus, tanh, sigmoid\}$. Among this plethora of possible neural networks we selected a specific subset as follows: we randomly selected three datasets slices of sizes $10^3, 10^4, 10^5$, and we ran a random search with 5-fold cross-validation on each of these three datasets; then we selected three neural networks which are capable of approximating short/medium/long sequences of input points.

For the choice of the breakpoint position $p$ at Line 14 of TOP-DOWN-REGRESSION, four strategies were tested: (i) midpoint $p = (a + b)/2$, which we used as baseline; (ii) random choice of $p \in (a, b)$; (iii) index of the highest error $p = a + \mathrm{argmax}_i errors[i]$, where *errors* is the array computed at Line 8; (iv) two breakpoints $p_1, p_2$ at the start/end positions of the longest chain of consecutive elements in *errors* that are greater than $\varepsilon$. On the Web logs dataset and $\varepsilon \in \{2^3, \dots, 2^8\}$, we found that with respect to the baseline (ii) produced on average 1.72% fewer models, (iii) produced 0.31% more models, and (iv) produced 11.08% fewer models. Therefore, we chose the strategy (iv) for the selection of the breakpoints at Line 14 of TOP-DOWN-REGRESSION.[10]

The Longitude dataset with $\varepsilon = 1024$ was covered with 2181 segments and 54 neural networks with four hidden units, thus a total of 2235 models. Restricting the pool to only segments, the algorithm produced instead 2254 segments, thus 19 more models. Perhaps surprisingly, the reduction in the number of models does not correspond to a reduction in space, since each neural network needed 13 parameters to be stored (in contrast to the two of a segment). The same phenomenon was observed on Web logs and IoT, and other $\varepsilon$ values. Probably the reason is that Top-Down-Regression assigns the responsibility of an area to the first model that does not make any error $> \varepsilon$, and it does not check whether the same area could have been covered by a more space efficient combination of simpler models. Doing this check naively would require a lengthy exhaustive search of all the combinations, so this issue deserves further research in the future.

---

[10] In terms of linear models, TOP-DOWN-REGRESSION is a greedy algorithm which produces on average 25.33% less linear models than the known greedy SC. In any case, we should also observe that it creates on average 48.61% more linear models than the optimal construction algorithm.

▶ **Summary.**  *The Hybrid PGM-index based on linear models and three very simple neural networks achieves up to 11% reduction in the number of models needed to $\varepsilon$-approximate the positions of the input keys. Nonetheless, the overall space occupancy is no better than the Recursive PGM-index based on linear models only. Despite these unsatisfying results, we think that the Hybrid PGM-index is very promising, but one needs to choose carefully the set of neural networks to select from (according to their offered trade-off between geometric power against space occupancy), and study techniques to build efficiently the PGM-index that suitably selects the best models from the set of available ones while avoiding the combinatorial explosion of an exhaustive search.*

## 5.2    Query performance of the PGM-index

We evaluated the average query performance of the PGM-index and other indexing data structures on Web logs, the biggest and most complex dataset available to us. The dataset was loaded in memory as a contiguous array of integers represented with 8 bytes and with 128 bytes payload. Slopes and intercepts were stored as double-precision floats. Each index was presented with 10M queries randomly generated on the fly.

**PGM-index variants.**  We experimented with the performance of the PGM-index using the three different indexing strategies described in Section 2.3: binary search, multiway tree (specifically we implemented a version of the CSS-tree [34]), and recursive geometric modelling. We refer to them, respectively, with PGM∘BIN, PGM∘CSS and PGM∘REC. We set $\varepsilon_\ell = 4$ for all but the last level of PGM∘REC, that is the one that includes the segments built over the input dataset by the optimal algorithm of the previous section.[11] Likewise, the node size of the CSS-tree was set to $B = 2\varepsilon_\ell$ for a fair comparison with the corresponding PGM∘REC.

Figure 11(left) shows that PGM∘REC dominates PGM∘CSS for $\varepsilon \leq 256$, and has better query performance than PGM∘BIN. The advantage of PGM∘REC over PGM∘CSS is also evident in terms of index height since the former has five levels whereas the latter has seven levels, thus a shorter traversal induced by a higher branching factor. For $\varepsilon > 256$ all the three variants behave similarly. The reason is that the PLA-model fits entirely into the L2 cache, hence there is no advantage in using an index on top of the models.

Finally, we considered larger values of $\varepsilon \geq 256$ and tested the performance of PGM∘BIN (recall that for those values all indexes behave similarly) with an exponential search that starts from the predicted position of the key in the leaf level. We found that exponential search is on average 16% slower than binary search for $256 \leq \varepsilon \leq 4096$, but it is 7% faster for

---

[11] The reason of this choice is that the number of segments is small, and recurring with a low $\varepsilon_\ell$ in such cases permits both a fast per-level search and a small space overhead for the index on top of the segments.

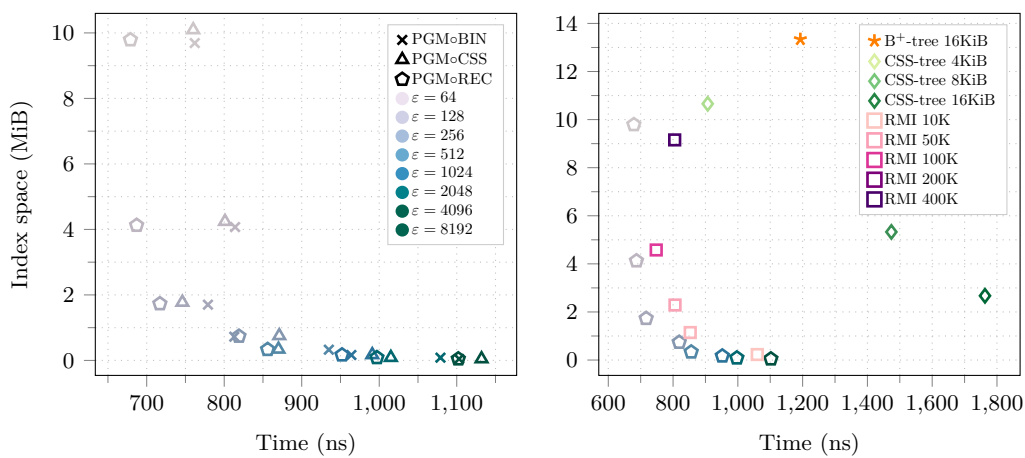

■ **Figure 11** On the left, the trade-offs offered by several configurations of PGM-index. On the right, the best configurations of PGM-index compared to the RMI with different second-stage sizes, and to traditional indexes with different page size, expressed in bytes.

$8192 \leq \varepsilon \leq 262144$. The reason is that the large jumps of the binary search are expensive in terms of I/Os when $\varepsilon$ is large, whereas the exponential search is more likely to have better I/O performances due to increased I/O locality.

**PGM-index vs traditional indexes.** We compared the PGM-index against two traditional indexes: the B+-tree and the CSS-tree [34]. For the former, we chose a well-known library [6], used as baseline also in [16, 27]. For the latter, we used our implementation.

The PGM-index dominated these traditional indexes, as shown in Figure 11(right) for page sizes in the range of 4–16 KiB. Performances for smaller page sizes are not plotted because they were too far from the main plot range. For example, the fastest CSS-tree in our machine had page size set to 128 bytes (twice the cache line), occupied 341 MiB and was matched in query performance by a PGM∘REC with $\varepsilon = 128$ which actually occupied only 4 MiB, thus inducing a reduction in space of 82.7×. As another example, the fastest B+-tree had page size set to 256 bytes, occupied 874 MiB and was matched in query performance by a PGM∘REC with $\varepsilon = 4096$ which actually occupied only 87 KiB, thus inducing a reduction in space of four orders of magnitude. It goes without saying that the B+-tree may also support insertions and deletions, which require a more flexible and space-hungry structure, so it is not a surprise that it performs worse than the CSS-tree and the PGM-index. Instead, what is surprising in this comparison, it is the amount of improvement in space occupancy achieved by the PGM-index which is four orders of magnitude with respect to B-trees and two orders of magnitude with respect to CSS-trees.

Overall, as stated in Section 1, traditional indexes are blind to the data distribution, and they miss the compression opportunities offered by its pattern and trends. On the contrary, adapting to the data distribution through linear approximations allows the PGM-index to uncover previously unknown space-time trade-offs, as we have demonstrated in this experiment.

**PGM-index vs RMI.** We implemented the 2-stage RMI of [27] using linear models and several numbers of models in the second stage.[12] Figure 11(right) shows that the PGM-index dominated RMI. The former has indeed better latency guarantees because, instead of fixing the structure beforehand and inspecting the errors afterwards, it is dynamically and optimally adapted to the input data distribution while guaranteeing the desired $\varepsilon$-approximation and trying to use the least possible space. Conversely, models in RMI are agnostic to the power of the models in the subsequent stages. Because of this, distributions of keys can be unbalanced in the last stage and can result in underused models. The most compelling evidence is the Mean Absolute Error (MAE) between the approximated and the predicted position, e.g., the PGM-index with $\varepsilon = 512$ needed ~32K segments and had MAE 226±139, while an RMI with the same number of second stage models (i.e. number of models at the last level) had MAE 892±3729 (3.9× more), thus higher and less predictable latency in the query execution. We expect this inefficiency of RMI to be even more severe when the data is stored on slower levels of the memory hierarchy, i.e., on disks or remote servers.

▶ **Summary.** *There is sufficient experimental evidence that the Recursive PGM-index is very flexible in trading query efficiency by compressed space occupancy, as demanded by big data applications. Moreover, the plots in Figure 11 show that the Recursive PGM-index dominates in time and space both the classic and the learned index structures. In particular, it improves the space occupancy of the CSS-tree by a factor 82.7× and the one of the B-tree by more than four orders of magnitude while achieving the same or even better query efficiency.*

## 5.3 Experimental performance of the Multicriteria PGM-index

Our implementation of the Multicriteria PGM-index operates in two modes: the time-minimisation mode (shortly, min-$t$) and the space-minimisation mode (min-$s$), which implement the algorithms described in Sections 4.1 and 4.2 respectively. In min-$t$ mode, inputs to the program are $s_{max}$ and a tolerance *tol* on the space occupancy of the solution, and the output is the value of $\varepsilon$ which guarantees a space bound $s_{max} \pm tol$. In min-$s$ mode, inputs to the program are $t_{max}$ and a tolerance *tol* on the time of the solution, and the output is the value of $\varepsilon$ which guarantees a time bound $t_{max} \pm tol$ in the query operations. We note that the introduction of a tolerance parameter allows us to stop the search earlier when any further step would not appreciably improve the solution (i.e., we seek only improvements of several bytes or nanoseconds). So *tol* is not a parameter that has to be tuned but rather a stopping criterion like the ones used in iterative methods. As further design choices we point out that: (i) the fitting of the power law, approximating $s(\varepsilon)$ or $t(\varepsilon)$, was performed with the Levenberg–Marquardt algorithm, while root finding was performed with Newton's method; (ii) the search space for $\varepsilon$ was set to $\mathcal{E} = [8, n/2]$ (since a cache line holds eight 64 bits integers); and finally (iii) the number of guesses was set to $2\lceil \log \log \mathcal{E} \rceil$.

**Experiments with the min-time mode.** Suppose that a database administrator wants the most efficient PGM-index for the Web logs dataset that fits into an L2 cache of 1 MiB. Our solver derived an optimal PGM-index matching that space bound by setting $\varepsilon = 393$ and taking 10 iterations and a total of 19 seconds. This result was obtained by approximating $s_L(\varepsilon)$ with the power law $46032135\,\varepsilon^{-1.16}$ which guarantees an error of no more than 4.8% over the range $\varepsilon \in [8, 1024]$. For comparison, the solver based on binary search took 27 iterations and 50 seconds.

---

[12] Personal correspondence with the authors clarified that the results in [27] were obtained with RMIs with linear models in both the stages, as we do in our experiments.

Now, suppose that a database administrator wants the most efficient PGM-index for the Longitude dataset that fits into an L1 cache of 32 KiB. Our solver derived an optimal PGM-index matching that space bound by setting $\varepsilon = 1050$ and taking 14 iterations and a total of 9 seconds. For comparison, the solver based on binary search took 20 iterations and 14 seconds.

**Experiments with the min-space mode.** Suppose that a database administrator wants the most compressed PGM-index for the IoT dataset that answers any query in less than 500 ns. Our solver derived an optimal PGM-index matching that time bound by setting $\varepsilon = 432$, occupying 74.55 KiB of space, and taking 9 iterations and a total of 6 seconds. For comparison, the solver based on binary search took 19 iterations and 12 seconds.

Now, suppose that a database administrator wants the most compressed PGM-index for the Web logs dataset that answers any query in less than 800 ns. Our solver derived an optimal PGM-index matching that time bound by setting $\varepsilon = 1217$, occupying 280.05 KiB of space, and taking 8 iterations and a total of 17 seconds. For comparison, the solver based on binary search took 24 iterations and 48 seconds.

▶ **Summary.** *There is sufficient experimental evidence that the multicriteria PGM-index effectively trades query efficiency by compressed space occupancy, as demanded by big data applications. With respect to the best known learned index, namely the RMI, which uses an expensive grid search, ours adopts a fast optimisation process that makes it suitable for applications with rapidly-changing data distributions and constraints.*

## 6 Conclusions and future work

In this paper, we introduced the PGM-index, a novel learned data structure for the dictionary problem. Compared to traditional data structures, our index is designed to trade smoothly, and more effectively than before, query time versus space occupancy by using a parameter $\varepsilon$ and the proper orchestration of geometric and learning concepts. We experimentally showed that it improves both query performance and space occupancy up to orders of magnitude, going well beyond classic indexes and modern learned indexes. Then, we extended its design to automatically deploy complex or simple regression models, and we tested its potential with neural networks. We introduced also a third variant of the PGM-index that adapts itself not only to the key distribution but also to the query distribution, still achieving very succinct space occupancy. Finally, we introduced the concept of Multicriteria Data Structures, showed that the PGM-index is an efficient and effective instance of this novel concept, and then designed an efficient solver that allows to specify a maximum query time and obtain the PGM-index that minimises the space, or vice versa. We demonstrated empirically that our solver halves the time to find the solution with respect to a baseline approach.

Numerous open problems and research opportunities need to be addressed:

**Hybrid indexes, other regression models and compressed tools.** We saw in Section 5.1.2 that when one employs nonlinear models, the procedure TOP-DOWN-REGRESSION can no longer guarantee that an area is covered by the PNA-model with the minimum possible space (see Section 2.2). Efficiently solving this problem is a challenging extension of our work and could open the way to designing very powerful hybrid indexes. Even more intriguing is the possibility of orchestrating segments, nonlinear models and techniques coming from the compression domain [1, 30, 40]. We believe that their use could be very suitable within our Multicriteria framework to design more powerful and effective hybrid learned indexes.

**A compressed variant of the PGM-index.** In this paper, we have focused only on the compactness of the index and assumed that the keys are stored uncompressed in a separate array $A$. In the future, we plan to study how to compress $A$ when it consists of integer keys while still being able to support fast membership, successor and predecessor queries. We remind the reader that indexing integer keys is of growing interest because of their occurrence in the posting lists of search engines (i.e. the docIDs) or in the adjacency lists of graphs (i.e. the nodeIDs, Web graphs or Social Network graphs) [39].

**The Data Calculator Engine.** We believe that the idea of building a PLA-model on a sequence of keys (Theorem 1) could become an effective design element for other data structures too. In this respect, we foresee its integration within the novel Data Calculator engine [22]. This could boost the design of new, or the re-design of classic, data structures.

**Handling deletions and insertions.** The incremental construction algorithm of the minimal-height strip shown in Section 2.1 allows managing insertions to the right of the current input keys. However, if the insertions can be arbitrary, then we could resort to split-merge operations over segments, like in classic B-trees, but this could possibly lose the minimality of the overall piecewise linear model (unless some overall reconstruction process is applied at fixed time instant given its high speed). The efficacy and closeness to minimality of this approach should be theoretically and experimentally investigated.

**Acknowledgements.**     We thank Carsten Binnig for providing us with the Web logs and IoT datasets. Tim Kraska for his advice on our implementation of his RMI. Gianna Del Corso for her helpful advice on the numerical algorithms implemented in our solver.